\documentclass{PoS}

\title{Singularities around the QCD critical point in the complex chemical potential plane}

\ShortTitle{Singularities around the QCD critical point}

\author{Shinji Ejiri\\
       Graduate School of Science and Technology, Niigata University,  Niigata 950-2181, Japan\\
        E-mail: \email{ejiri@muse.sc.niigata-u.ac.jp}}
        \author{Yasuhiko Shinno\\
       Nara National College of  Technology,Yamatokoriyama, Nara 639-1080, Japan\\
        E-mail: \email{shinno@libe.nara-k.ac.jp}}
\author{\speaker{Hiroshi Yoneyama}\\
        Department of Physics, Saga University, Saga 840-8502, Japan\\
        E-mail: \email{yoneyama@cc.saga-u.ac.jp}}

\abstract{We consider thermodynamic  singularities  appearing in the complex chemical potential  plane in the vicinity of QCD critical point.  In order to investigate 
what the singularities are like  in a concrete form,  we resort to an effective theory based on a mean filed approach.
  We   study  the behavior of  extrema of   the real part of the complex effective potential  in the complex order parameter  plane.  }

\FullConference{31st International Symposium on Lattice Field Theory - LATTICE 2013\\
		July 29 - August 3, 2013\\
		Mainz, Germany}

\begin{document}

\section{Introduction}
In this talk,    we report on  thermodynamic singularities of QCD 
  in the vicinity of the QCD critical point (CP)~\cite{AY, BCCGP, BR}   in the complex chemical potential ($\mu$)  plane. This issue is deeply connected  with zeros of the partition function. Study of them  provides an alternative approach to the critical phenomena and its scaling behaviors~\cite{YL,LY,IPZ,BMKKL,BMKKL2,ACFU,FK,DDLMZ,LM,E,NN}.  
  At finite volume, the zeros are scattered in the complex $\mu$ plane, and as the volume increases, the zeros accumulate 
  on  curves.    
  In the infinite volume limit, the zeros cross the real $\mu$ axis at the critical point  and the singularity appears on the real axis.  Off the critical point,  the singularities   moves  into the complex plane, and   the curve of zeros   called  the Stokes line~\cite{S} is  connected to the singularity.\par 
In order to investigate  what the  singularities are like in the vicinity of the CP,  we make use of   an effective theory~\cite{HI,FO} reflecting the phase structure around   the CP.  This model~\cite{HI} is constructed based on  the tricritical point (TCP) in the $\mu$-T plane in the chiral limit, which has the upper critical dimension equal to 3, and thus its  mean field description  is expected to be valid  up to a logarithmic correction.  Above the critical temperature,  the singularities are located off the real $\mu$ axis, and thus a reminiscence of the singularity appears on the real axis as a crossover. \par
With   complex $\mu$, the order parameter  becomes complex and so does the effective potential.   
Its $\mu$ dependence is quite intricate.  The simplicity of the   model, however,   allows to analyze the complex potential.  We explicitly trace  extrema of the real part of the potential.  
In the vicinity of the singularity,  their movements 
show different behaviors depending on where  they pass.  From its behaviors, we identify where the Stokes lines are located. \par
In the following section, we briefly explain  the model,  and the results are shown  in  section \ref{results}.   In  section \ref{summary},  summary is presented. 
\section{Model}
Let us briefly explain the model~\cite{HI} which we deal with in this report.
We consider two-flavor case. In the chiral  limit,  there exists a TCP  at finite temperature and density, which is connected to  a O(4)   critical point at $\mu=0$ along a critical line~\cite{PW}.   A first order line goes  down  from the TCP toward lower temperature side. 
When quark mass $m$ is introduced,  the critical line is absent, and the surviving first order line  terminates at a critical point  (CP). This   CP  can be described   by fluctuations of the  sigma meson  and is expected to share the same universality with 3-d Ising model~\cite{HJSSV}.  Since the upper critical dimension of the tricritical point is equal to 3,  the TCP in    QCD phase diagram  can  be described by a mean field theory up to a  logarithmic correction. 
As far as   $m$ is small,   the universal behavior around the CP is also expected to be described in the mean field framework.  \par
Starting with the Landau-Ginzburg potential 
\begin{eqnarray}
\Omega_{LG}=-m\sigma +\frac{a}{2}\sigma^2+\frac{b}{4}\sigma^4+\frac{c}{6}\sigma^6, 
\label{eq:TCP} 
\end{eqnarray}
one expands the coefficients  around the tricritical point (TCP) ($a=b=m=0$) as 
$
a(T,\mu)=C_a(T-T_3)+D_a(\mu-\mu_3), \
b(T,\mu)=C_b(T-T_3)+D_b(\mu-\mu_3),
$
where ($ \mu_3, T_3$) denotes  temperature and chemical potential at the TCP, respectively.  
The coefficients $C$ and  $D$ are constrained  to some extent reflecting  the  structure around the TCP,  i.e.,
$C_bD_a-C_aD_b>0 \ (C_a,C_b,D_a,D_b >0)$~\cite{HI}. 
 By switching on $m$, the CP exists at $T=T_E$ and $\mu=\mu_E$ so that 
$
 \Omega^{'}(T_{E}, \mu_{E}, \sigma_0)=\Omega^{''}(T_{E}, \mu_{E}, \sigma_0)=
 \Omega^{'''}(T_{E}, \mu_{E}, \sigma_0)=0 \ ( \Omega^{'}\equiv \partial \Omega/\partial \sigma), 
$
 which leads  to the coefficients 
$a(T_{E}, \mu_{E})=9b^2(T_{E}, \mu_{E})/(20 c)$,    and 
$-b(T_{E}, \mu_{E}) = (5/54^{1/5}) c^{3/5}m^{2/5}, $
and expectation value of $\sigma$; 
$
 \sigma_0=\sqrt{-3 \ b(T_{E}, \mu_{E})/(10 c)}.
$
Expanding $\Omega(T,\mu,\sigma)$ around  $\Omega(T_{E}, \mu_{E},\sigma_0)$, we obtain 
thermodynamic potential around the CP given by 
\begin{eqnarray}
\Omega(T,\mu,\sigma)=\Omega(T_{E}, \mu_{E}, \sigma_0)+A_1\hat {\sigma} +A_2\hat {\sigma}^2+A_3\hat {\sigma}^3+A_4\hat {\sigma}^4+\mathit{O}( {\hat\sigma}^5),
\label{eq:Omega_CEP}
\end{eqnarray}
where $\hat {\sigma}=\sigma-\sigma_0$.   
The coefficients $A_i$ are given as follows as  a function of $T$ and $\mu$.
\begin{eqnarray}
A_1&=&\left( C_a\sigma_0+C_b\sigma_0^3\right) \tilde{t}_E+\left( D_a\sigma_0+D_b \sigma_0^3\right)\tilde{\mu}_E\nonumber\\
A_2&=&\frac{1}{2}\left( C_a+3C_b\sigma_0^3\right) \tilde{t}_E +\frac{1}{2}\left( D_a+3D_b\sigma_0^3\right) \tilde{\mu}_E \nonumber\\
A_3&=& \left\{C_b \tilde{t}_E +D_b\tilde{\mu}_E \right\} \sigma_0\nonumber\\
A_4&=&-\frac{b(T_{E}, \mu_{E})}{2}+\frac{1}{4}\left(C_b\tilde{t}_E  +D_b\tilde{\mu}_E\right), 
\label{eq:coeff}
\end{eqnarray}
 where $\tilde{t}_E\equiv T-T_E$ and $\tilde{\mu}_E\equiv \mu-\mu_E$. 
\par
\section{Results}
 \label{results}
\subsection{Singular points}
Using the potential in Eq.~(\ref{eq:Omega_CEP}), 
 the instability of the extrema occurs at such $\sigma$  that
 \begin{eqnarray}
\frac{\partial \Omega}{\partial \sigma }=0, \quad \frac{\partial^2 \Omega}{\partial \sigma^2 }=0
\label{eq:singularity}
\end{eqnarray}
are simultaneously satisfied.   Namely, 
\begin{eqnarray}
A_1 +2 A_2\sigma+3 A_3\sigma^2+4 A_4\sigma^3=0, \quad 2 A_2+6 A_3\sigma+12 A_4\sigma^2=0.
\label{disc}
\end{eqnarray}
The discriminant of the left equation in Eq.(\ref{disc})  is solved as a function of $\tilde\mu_E$ and $\tilde t_E$ to obtain the singular points $\tilde{\mu}_E^{(s)}$.
%
 \begin{figure}
        \centerline{\includegraphics[width=14cm, height=6
cm]{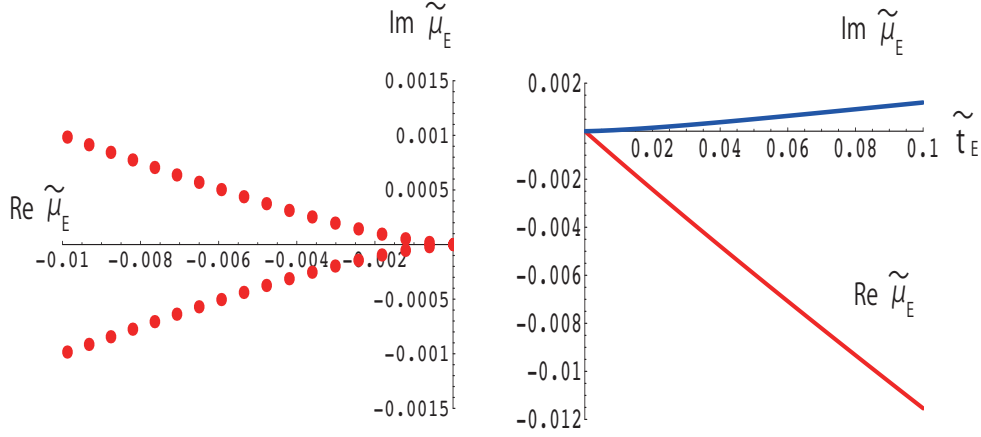}}
\caption{Left: Singularities in the complex $\tilde{\mu}_E=\mu-\mu_{E}$ plane. $C_a=0.1$ and $x_m=m^{1/5}=0.2$. The other parameters are $C_b=D_a=D_b=c=1.0$.  Temperatures are chosen to be  $\tilde{t}_E= T-T_{E}=0, 0.005, 0.01, 0.15, 0.02, \cdots,   0.09$.  Right:    The $\tilde{t}_E$-dependence of Re $\tilde{\mu}_E^{(s)}$ (red) and  Im $\tilde{\mu}_E^{(s)}$ (blue).  Re $\tilde{\mu}_E^{(s)}$ is linear in $\tilde{t}_E$, while  Im $\tilde{\mu}_E^{(s)}$ behaves  as $\tilde{t}_E^{3/2}$. 
 } 
\label{fig:edge-sing}
\end{figure}
The left panel of Fig.~\ref{fig:edge-sing} indicates the locations of  ${\tilde \mu}^{(s)}_E$ in the complex $\tilde{\mu}_E$  plane  for various values of temperature ($\tilde{t}_E\geq 0$).  Throughout  this report, $C_a=0.1$ and $C_b=D_a=D_b=c=1.0$ and  $x_m=m^{1/5}=0.2$ are chosen for  calculations. Temperatures are chosen to be  $\tilde{t}_E= T-T_{E}=0, 0.005, 0.01, 0.15, 0.02, \cdots,   0.09$.    At $\tilde{t}_E=0$, the singularity of the CP appears on the real $\tilde{\mu}_E$ axis (the origin in the figure).
For  $\tilde{t}_E> 0$,   the  singularities appearing in pairs deviate from the real $\tilde{\mu}_E$ axis as $\tilde{t}_E$ increases. The $\tilde{t}_E$-dependence of Re $\tilde{\mu}_E^{(s)}$ and Im  $\tilde{\mu}_E^{(s)}$ is shown in the right panel.   The real part depends  linearly on $\tilde{t}_E$, while the imaginary one behaves as $\tilde{t}_E^{3/2}$ as expected~\cite{S}.
\subsection{Extrema of the complex effective potential}
\label{sec:Stokes}
\begin{figure}
        \centerline{\includegraphics[width=10cm, height=5
cm]{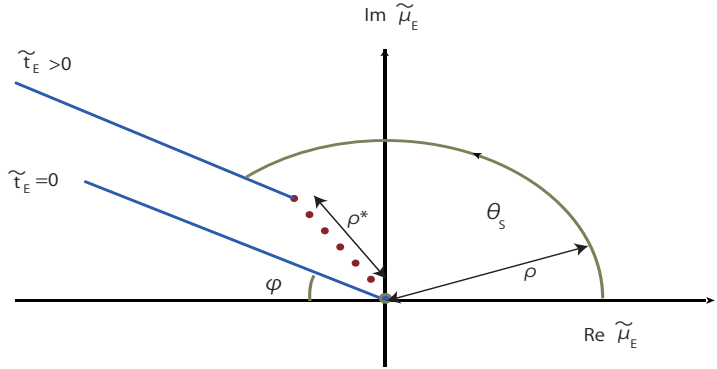}}
\caption{ Stokes lines (blue) in the complex   $\tilde{\mu}_E$ plane.  The singular point $\tilde{\mu}_E^{(s)}$ for  $\tilde{t}_E=0$ is located at the origin, while for  $\tilde{t}_E>0$ one of complex pair of $\tilde{\mu}_E^{(s)}$ is shifted to  the negative ${\rm Re } \tilde{\mu}_E$ direction in the upper half plane.   Angle $\varphi$ is defined from the negative side of ${\rm Re } \tilde{\mu}_E$ axis.   }
  \label{fig:Stokes-1}
\end{figure}
  In the present system, 
the critical point   exists at    $\tilde{\mu}_E=0$ for  $\tilde{t}_E=0$,   
while for  $\tilde{t}_E >0$ the singular point  is shifted into the complex plane $\tilde{\mu}_E^{(s)}=\rho^{(s)} e^{i \theta^{(s)}}$ ($\rho^{(s)}>0$)  as shown in  Fig.~\ref{fig:Stokes-1}, where only the upper half plane is shown.  
Let  us   move  $\tilde{\mu}_E$   from a point on the real axis ($\tilde{\mu}_E>0$) 
 to the other side  at $\tilde{\mu}_E<0$  by changing $\theta$ from 0 to $\pi$ in $\tilde{\mu}_E=\rho e^{i \theta}$ for a fixed value of $\rho$.  As the parameter is   analytically continued from the one phase to the other, the corresponding global minimum of the potential  is also analytically continued.  \par
\begin{figure}
        \centerline{\includegraphics[width=14cm, height=6
cm]{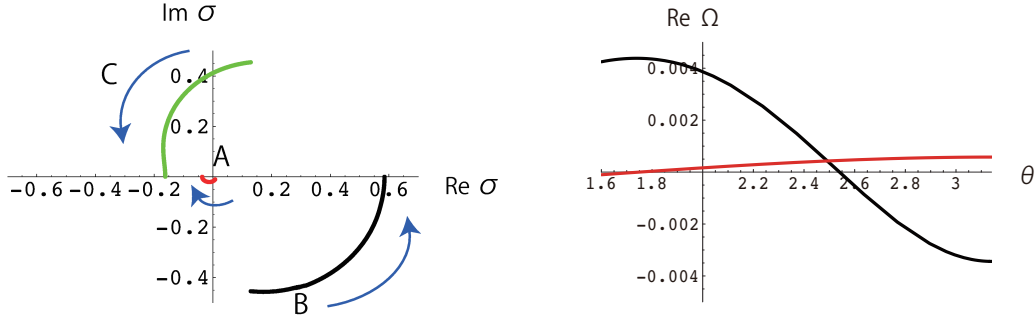}}
\caption{ Left: Behaviors of   the positions of three extrema  of ${\rm Re}\ \Omega$ in the complex $\sigma$ plane   as   $\tilde{\mu}_E=\rho e^{i \theta}$ varies from  $\theta=0$ to $\theta=\pi$ ($\rho=0.05$) for $\tilde{t}_E=0.1$.    Right: Corresponding behaviors of  Re\ $\Omega$ for the two extrema A (red) and B (black) as a function of $\theta$. The two values agree at $\theta_S=2.492$. 
 }
\label{fig:extrema1}
\end{figure}

 In order to explicitly see how this occurs, we take  $\tilde{t}_E=0.1$.  At this temperature,   the singular points  are located  at $\tilde{\mu}_E^{(s)}=-0.0115 \pm i \ 0.0012$ and thus $\rho^{(s)}=0.0116$. The left panel of 
Fig.~\ref{fig:extrema1} shows behaviors of   the positions of three extrema  of ${\rm Re}\ \Omega$ in the complex $\sigma$ plane as  $\theta$ varies from  $0$ to $\pi$ with a fixed value of $\rho$ $(=0.05)$.   
 For $\theta=0$, ${\rm Re}\ \Omega$  develops three   extrema  at $\sigma=0.0108$ (A) 
and  $\sigma=0.12910-i \ 0.4544$  (B) and $0.12910+i\  0.4544$ (C), each of which is  located at the initial point of the arrows denoted by A, B and C, respectively.   
The arrows in the figure indicate the direction of the movement  of the three as $\theta$ varies from 0 to $\pi$. 
The extremum  A, which is a global minimum for $\theta=0$,  ends up with  negative small  value of  $\sigma=-0.0366$ at $\theta=\pi$, while 
 B does  with the real axis at  $\sigma=0.5848$,  which is identified as the  global minimum of $\Omega$ for $\theta=\pi$ from its shape.  
  So the two extrema  A and B are associated with the global minimum of  $\Omega$  for real $\mu$, the phase for  $\tilde{\mu}_E>0$   and the  one for $\tilde{\mu}_E<0$ , respectively, while the  extremum C  is not associated  with the phase transition.  In this case, a discrete change occurs from one minimum to the other  between $\theta=0$  and  $\pi$, where 
 \begin{eqnarray}
{\rm Re}\  \Omega(\sigma_+)&=&{\rm Re}\  \Omega(\sigma_-),
 \label{eq2}
\end{eqnarray}
is fulfilled,  
where $\sigma_+ (\sigma_-)$ is the location of the extremum analytically continued from $\tilde{\mu}_E>0$ ($\tilde{\mu}_E<0$ ). For $\rho=0.05$,  it occurs at
at $\theta=\theta_S=2.492=0.793\ \pi$ as shown in the right panel of Fig.~\ref{fig:extrema1}, where their slopes show a discontinuity  in accordance with the interpretation 
that the distribution of the partition function zeros can be regarded  as that of the charges
in  analogy with two dimensional Coulomb gas~\cite{YL,S}.  \par
As  $\rho$ decreases, the two extrema A and B approach each other, and for   $\rho=\rho^{(s)}$,  the two trajectories meet together at $\sigma^{(s)}$, where   $\sigma^{(s)}$ is   the minimum of  ${\rm Re}\ \Omega$ for $\tilde{\mu}_E$ corresponding to the singularity, $\tilde{\mu}_E^{(s)}$ i.e.,  fulfilling the condition Eq. (\ref{eq:singularity}).  In the case of  $\tilde t_E=0.1$, $ \tilde{\mu}_E^{(s)}=-0.0115 +i \ 0.0012$, and  $\sigma^{(s)}=0.1219- i \ 0.0653$.  
 For   further smaller value of $\rho<\rho^{(s)}$,  only a single extremum A is associated with analytic continuation from $\tilde{\mu}_E>0$ to $\tilde{\mu}_E<0$.
   The other two extrema do not play a role for vacua.

\subsection{Stokes lines}
\begin{figure}
        \centerline{\includegraphics[width=10cm, height=6
cm]{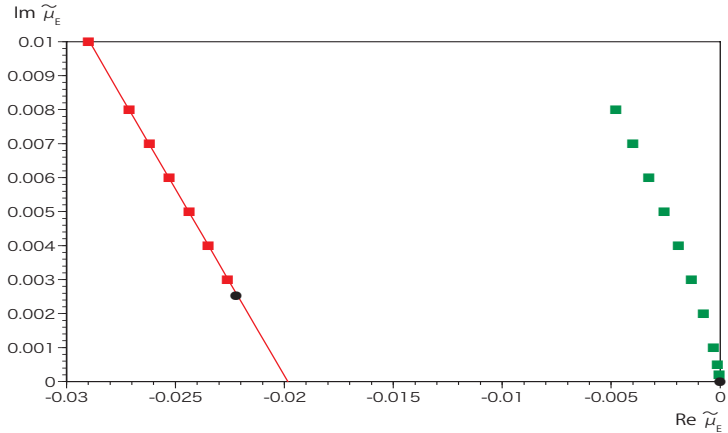}}
\caption{ Locations of the Stokes line in complex $\tilde{\mu}_E$ plane, which are calculated   by the trajectories of  the two extrema of  ${\rm Re}\ \Omega$   for $\tilde{t}_E=0.2$  (red)  and $\tilde{t}_E=0$  (green).      For  $\tilde{t}_E=0.2$, it is clearly seen that thus identified points sit on a straight line with the slope $-1.09\pm 0.02$. Black symbol denotes the locations of the singular points $\tilde{\mu}_E^{(s)}$ $=-0.02221+i \ 0.00254$. For  $\tilde{t}_E=0$, the singularity corresponding to the CP is situated at the origin.  In this case the Stokes line emanates from the origin with $\varphi=\pi/2$.  
}
\label{fig:Stokes-ttil00-02}
\end{figure}

In the case of $\tilde{t}_E>0$, therefore, the Stokes line runs for $\rho>\rho^{(s)}$.  
Figure~\ref{fig:Stokes-ttil00-02} indicates how the Stokes line emanates from the singular point $\tilde{\mu}_E^{(s)}$ in the complex   $\tilde{\mu}_E$ plane. Squared  symbols indicate the locations which are  calculated in the way  described above, i.e., by choosing several value of $\rho$, tracing  trajectories of the positions of the extrema upon varying $\theta$ and  checking the behaviors of Re $\Omega$.   
   For  $\tilde{t}_E=0.2$,  the singular point is located at $\tilde{\mu}_E^{(s)}$$=-0.02221+i\ 0.00254$ (black filled  circle), and  the  Stokes line is alined on a straight line, which   is  tilted with  angle   $\varphi=\pi/4$ from  an axis parallel to the  negative axis of ${\rm Re} \tilde{\mu}_E$.  
 As $\tilde{t}_E$ decreases, $\tilde{\mu}_E^{(s)}$ approaches the origin, and  the angle    $\varphi$   shows a increasing tendency.      
 Green   filled squares  show  the locations of the Stokes line  
  for $\tilde{t}_E=0$. In this case the Stokes line emanates from the origin with $\varphi=\pi/2$.   \par
  For   $\tilde{t}_E=-0.2$, 
 a first order phase transition is located at $\tilde{\mu}_E=0.05553$,   and  
 the  Stokes line goes out  upright with $\varphi=\pi/2$ .  In recent Monte Carlo study~\cite{NMNNS} of low temperature and high density QCD,   the distribution of  the Lee-Yang zeros have been calculated, and it   looks similar to the  behavior in the case of $\tilde{t}_E<0$ ($\varphi=\pi/2$),  suggesting  a possible first order phase transition. 
\section {Summary}
\label{summary}
We have discussed  the thermodynamic singularities in the complex chemical potential plane in  QCD  at  finite temperature and finite densities. For  this  purpose, we adopted an effective theory incorporating fluctuations around  the CP.       Singularities  in the complex chemical potential plane are identified as    unstable points  of the  extrema  of the complex effective  potential. 
At  CP temperature, the singularity is located on the real $\mu$ axis, and   above CP temperature  it  moves away from the real axis leaving its  reminiscence   as a   crossover.  Some  phenomena on the real axis  are then  deeply connected with   the singularity  associated with the CP.  Study of this  aspect will be reported elsewhere~\cite{ESY}.  
Simplicity  of the model allows us to explicitly deal with the complex potential as a function of the complex order parameter and complex $\mu$. 
  We have had  a close look at the behavior of  the extrema of ${\rm Re}\ \Omega(\sigma)$ in the complex order parameter plane.   It is found  that  two relevant extrema make a rearrangement at the singular point under the variation of $\mu$ around the singularity in the complex $\mu$ plane.     
It is also explicitly  seen that  the Stokes lines are  located in the different way depending on temperature,  higher or  lower than the  CP temperature or at the  CP temperature.  This  provides us with  information as to where the Lee-Yang zeros are located for finite volume.  
 \par
It may  be  worth while   to quantify what is described here.  
  In \cite{EY},   singularities of  two-flavor QCD with staggered quarks have been investigated   by having a look at  the effective potential with respect to the plaquette variable.   In order to have  a more contact with the present report,  some refinement of the computation would be necessary.  This is now in progress~ \cite{EY2}. 

 \section*{Acknowledgment}
S. E.  is in part supported by Grants-in-Aid of the Japanese Ministry of Education, Culture, Sports, Science and Technology (No. 23540259).

\end{document}